\documentclass{article}
\usepackage{graphicx}

\begin{document}

\author{G. Quznetsov \\
Chelyabinsk State University\\
Chelyabinsk, Ural, Russia\\
quznets@yahoo.com, gunn@mail.ru}
\title{Double-Slit Experiment and Quantum Theory Event-Probability Interpretation}
\date{February 18, 2010}
\maketitle

\begin{abstract}
In this article the propagation of pointlike event probabilities in space is considered. 
Double-Slit experiment is described in detail. New interpretation of Quantum Theory is formulated.
\end{abstract}

\section{Introduction}
\label{intro}

\rule{12pt}{0pt} Ontology and interpretation of Quantum Mechanics are discussed from the thirties of the XX 
century \cite{EPR} till present days \cite{arx}. A clear description of these basic interpretations is presented 
in the book by Anthony Sudbery \cite{Anthony}.

 I present another interpretation of Quantum Mechanics. This interpretation is based on 
newly-discovered fact that probabilities of pointlike events can be expressed by complex 4X1 matrix functions. 
And these functions obey equations which are similar to the Dirac's equations \cite{PP}.

 And here I evolve the idea of H. Bergson, A. N. Whitehead, M. Capek, E. C. Whipple jr., J. Jeans 
of presentation of elementary particles by events \cite{WWW}.

\section{Propagation of Probability in Space}
\label{sec:1}

\rule{12pt}{0pt} Let\footnote{Denote: $\underline{x}:=\left\langle x_0,\mathbf{x}\right\rangle :=\left\langle
x_0,x_1,x_2,x_3\right\rangle $,
$t:=\left( 1/\mathrm{c}\right) \cdot x_0$, c = 299792458.}  $\rho _{\mathtt{A}}\left( \underline{x}\right) $ be a probability
density of a point event $\mathtt{A}\left( \underline{x}\right) $. And let
real functions

\[
u_{\mathtt{A},1}\left( \underline{x}\right) ,u_{\mathtt{A},2}\left( 
\underline{x}\right) ,u_{\mathtt{A},3}\left( \underline{x}\right) 
\]

satisfy conditions

\[
u_{\mathtt{A},1}^2+u_{\mathtt{A},2}^2+u_{\mathtt{A},3}^2<\mathrm{c}^2%
\mbox{,}
\]

and let if $j_{\mathtt{A},s}:=\rho _{\mathtt{A}}u_{\mathtt{A},s}$ then

\begin{eqnarray*}
&&\displaystyle \rho _{\mathtt{A}}\rightarrow \rho _{\mathtt{A}}^{\prime }=\frac{\rho _{%
\mathtt{A}}-\frac v{\mathrm{c}^2}j_{\mathtt{A},k}}{\sqrt{1-\left( \frac v{%
\mathrm{c}}\right) ^2}}\mbox{,} \\
&&\displaystyle j_{\mathtt{A},k}\rightarrow j_{\mathtt{A},k}^{\prime }=\frac{j_{\mathtt{%
A},k}-v\rho _{\mathtt{A}}}{\sqrt{1-\left( \frac v{\mathrm{c}}\right) ^2}}%
\mbox{,} \\
&&\displaystyle j_{\mathtt{A},s}\rightarrow j_{\mathtt{A},s}^{\prime }=j_{\mathtt{A},s}%
\mbox{ for }s\neq k
\end{eqnarray*}

for $s\in \left\{ 1,2,3\right\} $ and $k\in \left\{ 1,2,3\right\} $ under
the Lorentz transformations:

\begin{eqnarray*}
\displaystyle t &\rightarrow &t^{\prime }=\frac{t-\frac v{\mathrm{c}^2}x_k}{\sqrt{1-\frac{%
v^2}{\mathrm{c}^2}}}\mbox{, } \\
\displaystyle x_k &\rightarrow &x_k^{\prime }=\frac{x_k-vt}{\sqrt{1-\frac{v^2}{\mathrm{c}^2%
}}}\mbox{, } \\
x_s &\rightarrow &x_s^{\prime }=x_s\mbox{,
if }s\neq k\mbox{.}
\end{eqnarray*}

\rule{12pt}{0pt} In that case $\mathbf{u}_{\mathtt{A}}\left\langle u_{\mathtt{A},1},u_{%
\mathtt{A},2},u_{\mathtt{A},3}\right\rangle $ is called \textit{a vector
of local velocity} of an event $\mathtt{A}$ \textit{probability propagation}
and

\[
\mathbf{j}_{\mathtt{A}}\left\langle j_{\mathtt{A},1},j_{\mathtt{A}%
,2},j_{\mathtt{A},3}\right\rangle 
\]

is called \textit{a current vector} of an  event $\mathtt{A}$ probability.

\rule{12pt}{0pt} Let us consider the following set of four real equations with eight real
unknowns:

\[
b^2 \mbox{ with } b>0\mbox{, } {\alpha }\mbox{, } {\beta }\mbox{, } {\chi }\mbox{, } {\theta }\mbox{, } 
{\gamma }\mbox{, } {\upsilon }\mbox{, } {\lambda }\mbox{:}
\]

\vspace*{2pt}
\begin{equation}
\rule{-0.5cm}{0pt}\left\{
\begin{array}{l}
\displaystyle b^2=\rho _{\mathtt{A}}\mbox{,}  \\[11pt]
\displaystyle b^2\left( 
\begin{array}{l}
\cos ^2\left( {\alpha }\right) \sin \left( 2{\beta }\right) \cos \left( {%
\theta }-{\gamma }\right)  \\ 
-\sin ^2\left( {\alpha }\right) \sin \left( 2{\chi }\right) \cos \left( {%
\upsilon }-{\lambda }\right) 
\end{array}
\right) =-\frac{j_{\mathtt{A},1}}{\mathrm{c}}\mbox{,}  \\[11pt]
\displaystyle b^2\left( 
\begin{array}{l}
\cos ^2\left( {\alpha }\right) \sin \left( 2{\beta }\right) \sin \left( {%
\theta }-{\gamma }\right)  \\ 
-\sin ^2\left( {\alpha }\right) \sin \left( 2{\chi }\right) \sin \left( {%
\upsilon }-{\lambda }\right) 
\end{array}
\right) =-\frac{j_{\mathtt{A},2}}{\mathrm{c}}\mbox{,}  \\[11pt]
\displaystyle b^2\left( 
\begin{array}{l}
\cos ^2\left( {\alpha }\right) \cos \left( 2{\beta }\right)  \\ 
-\sin ^2\left( {\alpha }\right) \cos \left( 2{\chi }\right) 
\end{array}
\right) =-\frac{j_{\mathtt{A},3}}{\mathrm{c}}\mbox{.}
\end{array}
\right.
\label{abc}
\end{equation}

\rule{12pt}{0pt} This set has solutions for any $\rho _{\mathtt{A}}$ and $j_{\mathtt{A}%
,k}$. For example, one of these solutions can be found in \cite{PTGT}.

\rule{12pt}{0pt} If
 
\begin{eqnarray}
&&\varphi _1 :=b\cdot\exp \left( \mathrm{i}{\gamma }\right) \cos \left( {\beta }%
\right) \cos \left( {\alpha }\right) \mbox{,}  \nonumber \\
&&\varphi _2 :=b\cdot\exp \left( \mathrm{i}{\theta }\right) \sin \left( {\beta }%
\right) \cos \left( {\alpha }\right) \mbox{,}  \nonumber \\
&&\varphi _3 :=b\cdot\exp \left( \mathrm{i}{\lambda }\right) \cos \left( {\chi }%
\right) \sin \left( {\alpha }\right) \mbox{,}  \label{xxx55} \\
&&\varphi _4 :=b\cdot\exp \left( \mathrm{i}{\upsilon }\right) \sin \left( {\chi }%
\right) \sin \left( {\alpha }\right)   \nonumber
\end{eqnarray}

then 

\begin{eqnarray}
\rho _{\mathtt{A}} &=&\sum_{s=1}^4\varphi _s^{*}\varphi _s\mbox{,}
\label{j} \\
\frac{j_{\mathtt{A},r }}{\mathrm{c}} &=&-\sum_{k=1}^4\sum_{s=1}^4%
\varphi _s^{*}\beta _{s,k}^{\left[r \right] }\varphi _k  \nonumber
\end{eqnarray}

with $r\in \left\{ 1,2,3\right\} $ and with

\begin{eqnarray*}
\beta ^{[1]} &:&=\left[ 
\begin{array}{cccc}
0 & 1 & 0 & 0 \\ 
1 & 0 & 0 & 0 \\ 
0 & 0 & 0 & -1 \\ 
0 & 0 & -1 & 0
\end{array}
\right] ,\beta ^{[2]}:=\left[ 
\begin{array}{cccc}
0 & -\mathrm{i} & 0 & 0 \\ 
\mathrm{i} & 0 & 0 & 0 \\ 
0 & 0 & 0 & \mathrm{i} \\ 
0 & 0 & -\mathrm{i} & 0
\end{array}
\right]  \\
,\beta ^{[3]} &:&=\left[ 
\begin{array}{cccc}
1 & 0 & 0 & 0 \\ 
0 & -1 & 0 & 0 \\ 
0 & 0 & -1 & 0 \\ 
0 & 0 & 0 & 1
\end{array}
\right] .
\end{eqnarray*}

\rule{12pt}{0pt} These functions $\varphi _s$ are called \textit{functions of event} $\mathtt{%
A}$ \textit{state}.

\rule{12pt}{0pt} If $\rho _{\mathtt{A}}\left(\underline{x}\right) =0$ for all $\underline{x}$
such that $\left|\underline{x}\right| >\left( \pi \mathrm{c} /\mathrm{h}\right)$ with 
$\mathrm{h}:=$ $6.6260755\cdot 10^{-34}$ then $\varphi _s\left(\underline{x}%
\right) $ are Planck's functions \cite{PP}. And if 
\[
\varphi :=\left[ 
\begin{array}{c}
\varphi _1 \\ 
\varphi _2 \\ 
\varphi _3 \\ 
\varphi _4
\end{array}
\right] 
\]

then there exist matrix $\widehat{Q}$ so that 

\begin{equation}
\widehat{Q}=\left[ 
\begin{array}{cccc}
\mathrm{i}\vartheta _{1,1} & \mathrm{i}\vartheta _{1,2}-\varpi _{1,2} & 
\mathrm{i}\vartheta _{1,3}-\varpi _{1,3} & \mathrm{i}\vartheta _{1,4}-\varpi
_{1,4} \\ 
\mathrm{i}\vartheta _{1,2}+\varpi _{1,2} & \mathrm{i}\vartheta _{2,2} & 
\mathrm{i}\vartheta _{2,3}-\varpi _{2,3} & \mathrm{i}\vartheta _{2,4}-\varpi
_{2,4} \\ 
\mathrm{i}\vartheta _{1,3}+\varpi _{1,3} & \mathrm{i}\vartheta _{2,3}+\varpi
_{2,3} & \mathrm{i}\vartheta _{3,3} & \mathrm{i}\vartheta _{3,4}-\varpi
_{3,4} \\ 
\mathrm{i}\vartheta _{1,4}+\varpi _{1,4} & \mathrm{i}\vartheta _{2,4}+\varpi
_{2,4} & \mathrm{i}\vartheta _{3,4}+\varpi _{3,4} & \mathrm{i}\vartheta
_{4,4}
\end{array}
\right]   \label{Q}
\end{equation}

with real $\varpi _{s,k}$ and $\vartheta _{s,k}$ and $\varphi $ obeys the following 
differential equation \cite{PP}: 
\begin{equation}
\partial _t\varphi =\mathrm{c}\left( \beta ^{\left[ 1\right] }\partial
_1+\beta ^{\left[ 2\right] }\partial _2+\beta ^{\left[ 3\right] }\partial _3+%
\widehat{Q}\right) \varphi   \label{ham1}
\end{equation}

\rule{12pt}{0pt} In that case if
 
\[
\widehat{H}=\mathrm{ic}\left( \beta ^{\left[ 1\right] }\partial _1+\beta
^{\left[ 2\right] }\partial _2+\beta ^{\left[ 3\right] }\partial _3+\widehat{Q}\right) %
\]

then $\widehat{H}$ is called \textit{a Hamiltonian} of a moving with
equation (\ref{ham1}).

\rule{12pt}{0pt} Operator $\widehat{U}\left( t,t_0\right) $ with domain and with range
of values on the set of state vectors is called {\it an evolution
operator} if each state vector $\varphi $ fulfills the following condition:

\begin{equation}
\varphi \left( t\right) =\widehat{U}\left( t,t_0\right) \varphi \left(
t_0\right) \mbox{.}  \label{U100}
\end{equation}

\rule{12pt}{0pt} Let us denote:

\[
\widehat{H}_d:=\mathrm{c}\sum_{s=1}^3\mathrm{i}\beta ^{\left[
s\right] }\partial _s\mbox{.} 
\]

\rule{12pt}{0pt} In that case

\[
\widehat{H}=\widehat{H}_d+\mathrm{ic}\widehat{Q} 
\]

according to the Hamiltonian definition.

\rule{12pt}{0pt} From (\ref{ham1}):

\[
\mathrm{i}\partial _t\varphi =\widehat{H}\varphi \mbox{.} 
\]

\rule{12pt}{0pt} Hence,

\[
\mathrm{i}\partial _t\varphi =\left( \widehat{H}_d+\mathrm{ic}\widehat{Q}%
\right) \varphi \mbox{.} 
\]

\rule{12pt}{0pt} This differential equation has got the following solution:

\[
\varphi \left( t\right) = \left( \exp \left( -\mathrm{i}%
\widehat{H}_d\left( t-t_0\right) +\mathrm{c}\int_{t=t_0}^t\widehat{Q}%
\partial t\right)\right) \varphi \left( t_0\right) \mbox{.} 
\]

\rule{12pt}{0pt} Hence, from (\ref{U100}):

\[
\widehat{U}\left( t,t_0\right) =\exp \left( -\mathrm{i}\widehat{H}_d\left(
t-t_0\right) +\mathrm{c}\int_{t=t_0}^t\widehat{Q}\partial t\right) 
\]

\rule{12pt}{0pt} Fourier series for $\varphi _j\left( t,\mathbf{x}\right) $ has the following shape \cite{PP}:

\[
\varphi _j\left( t_0,\mathbf{x}\right) =\sum_{\mathbf{p}}c_{j,\mathbf{p}%
}\left( t_0\right) \mathbf{\varsigma }_{\mathbf{p}}\left( t_0,\mathbf{x}%
\right) 
\]

with

\[
\mathbf{\varsigma }_{\mathbf{p}}\left( \mathbf{x}\right) :=\left\{ 
\begin{array}{c}
\left( \frac{\mathrm{h}}{2\pi \mathrm{c}}\right) ^{\frac 32}\exp \left( -%
\mathrm{i}\frac{\mathrm{h}}{\mathrm{c}}\mathbf{px}\right) \mbox{ if } \\ 
\frac{-\pi \mathrm{c}}{\mathrm{h}}\leq x_k\leq \frac{\pi \mathrm{c}}{\mathrm{%
h}}\mbox{ for }k\in \left\{ 1,2,3\right\} \mbox{;} \\ 
0\mbox{, otherwise}
\end{array}
\right| \mbox{.}
\]

\rule{12pt}{0pt} Therefore, in accordance with properties of Fourier's transformation:

\begin{eqnarray*}
&&\varphi \left( t,\mathbf{x}\right) = \\
&&\int\limits_{-\frac{\pi \mathrm{c}}{\mathrm{h}}}^{\frac{\pi \mathrm{c}}{%
\mathrm{h}}}\int\limits_{-\frac{\pi \mathrm{c}}{\mathrm{h}}}^{\frac{\pi 
\mathrm{c}}{\mathrm{h}}}\int\limits_{-\frac{\pi \mathrm{c}}{\mathrm{h}}}^{%
\frac{\pi \mathrm{c}}{\mathrm{h}}}d\mathbf{x}_0\cdot \left( \frac{\mathrm{h}%
}{2\pi \mathrm{c}}\right) ^3\left( 
\begin{array}{c}
\sum_{\mathbf{p}}\exp \left( -\mathrm{i}\widehat{H}_d\left( t-t_0\right) +%
\mathrm{c}\int_{t=t_0}^t\widehat{Q}\partial t\right)\times  \\ 
\times \exp \left( -\mathrm{i}\frac{\mathrm{h}}{\mathrm{c}}\mathbf{p}\left( 
\mathbf{x-x}_0\right) \right) 
\end{array}
\right)\times  \\
&&\times \varphi \left( t_0,\mathbf{x}_0\right) \mbox{.}
\end{eqnarray*}

\rule{12pt}{0pt} An operator

\begin{eqnarray*}
&&K\left( t-t_0,\mathbf{x-x}_0,t,t_0\right):=  \\
&&\left( \frac{\mathrm{h}}{2\pi \mathrm{c}}\right) ^3\left( 
\begin{array}{c}
\sum_{\mathbf{p}}\exp \left( -\mathrm{i}\widehat{H}_d\left( t-t_0\right) +%
\mathrm{c}\int_{t=t_0}^t\widehat{Q}\partial t\right)\times  \\ 
\times \exp \left( -\mathrm{i}\frac{\mathrm{h}}{\mathrm{c}}\mathbf{p}\left( 
\mathbf{x-x}_0\right) \right) 
\end{array}
\right) 
\end{eqnarray*}

is called {\it a propagator } of the  $\mathcal{A}$ probability.

\rule{12pt}{0pt} Hence,

\begin{equation}
\varphi \left( t,\mathbf{x}\right) =\int\limits_{-\frac{\pi \mathrm{c}}{%
\mathrm{h}}}^{\frac{\pi \mathrm{c}}{\mathrm{h}}}\int\limits_{-\frac{\pi 
\mathrm{c}}{\mathrm{h}}}^{\frac{\pi \mathrm{c}}{\mathrm{h}}}\int\limits_{-%
\frac{\pi \mathrm{c}}{\mathrm{h}}}^{\frac{\pi \mathrm{c}}{\mathrm{h}}}d%
\mathbf{x}_0\cdot K\left( t-t_0,\mathbf{x-x}_0,t,t_0\right) \varphi \left(
t_0,\mathbf{x}_0\right) \mbox{.}  \label{prp}
\end{equation}

\rule{12pt}{0pt} {\bf But this propagator acts for the probability, but not for particles.}

\rule{12pt}{0pt} A propagator has the following property:

\begin{eqnarray*}
&&K\left( t-t_0,\mathbf{x-x}_0,t,t_0\right) = \\
&&\int\limits_{-\frac{\pi \mathrm{c}}{\mathrm{h}}}^{\frac{\pi \mathrm{c}}{%
\mathrm{h}}}\int\limits_{-\frac{\pi \mathrm{c}}{\mathrm{h}}}^{\frac{\pi 
\mathrm{c}}{\mathrm{h}}}\int\limits_{-\frac{\pi \mathrm{c}}{\mathrm{h}}}^{%
\frac{\pi \mathrm{c}}{\mathrm{h}}}d\mathbf{x}_1\cdot K\left( t-t_1,\mathbf{%
x-x}_1,t,t_1\right) \times  \\
&&\times K\left( t_1-t_0,\mathbf{x}_1\mathbf{-x}_0,t_1,t_0\right) \mbox{.}
\end{eqnarray*}

\section{Double-Slit Experiment}

\begin{figure}
\includegraphics[width=0.7\textwidth]{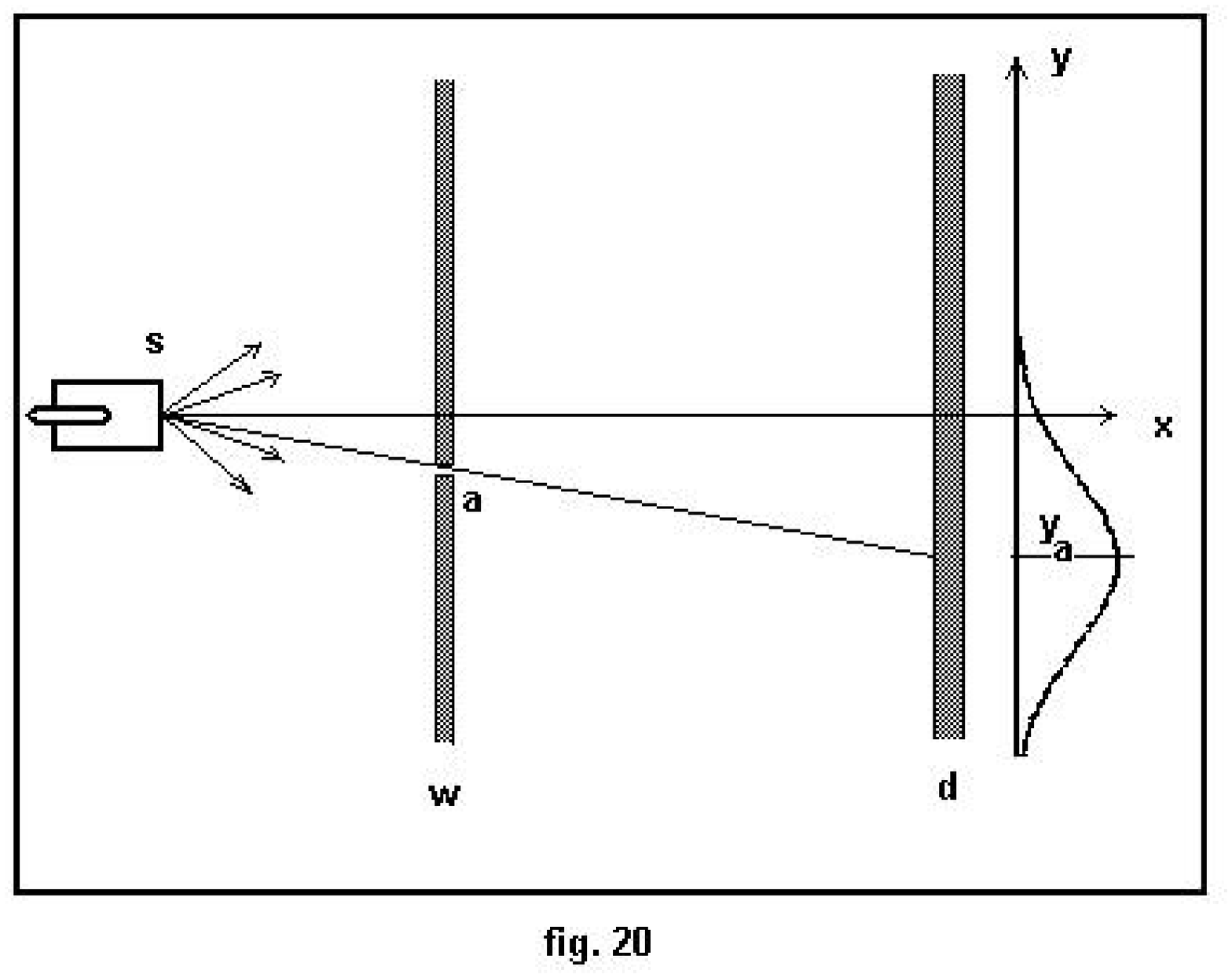}
\caption{}
\label{fig:20}
\end{figure}

\begin{figure}
\includegraphics[width=0.7\textwidth]{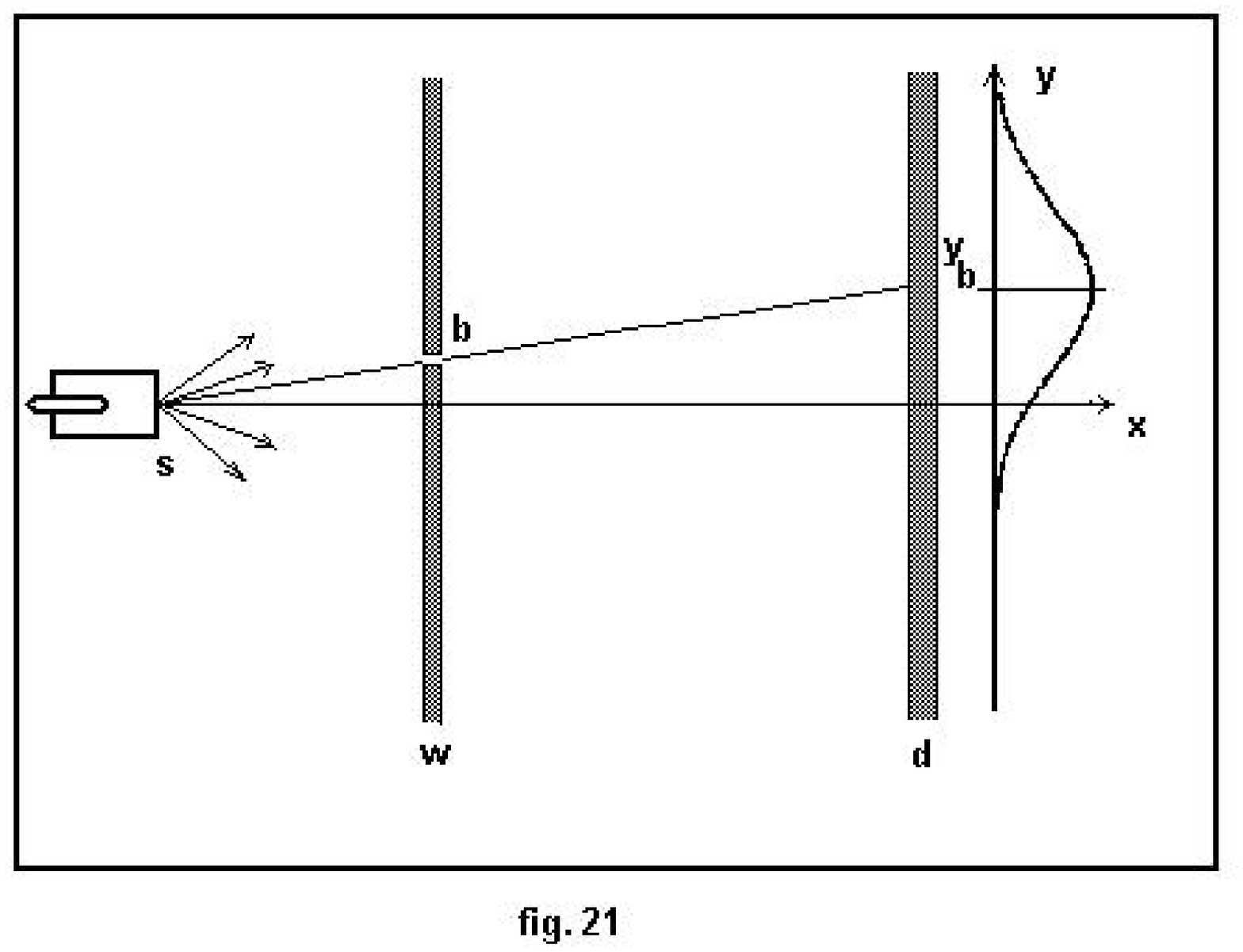}
\caption{}
\label{fig:21}
\end{figure}

\begin{figure}
\includegraphics[width=0.7\textwidth]{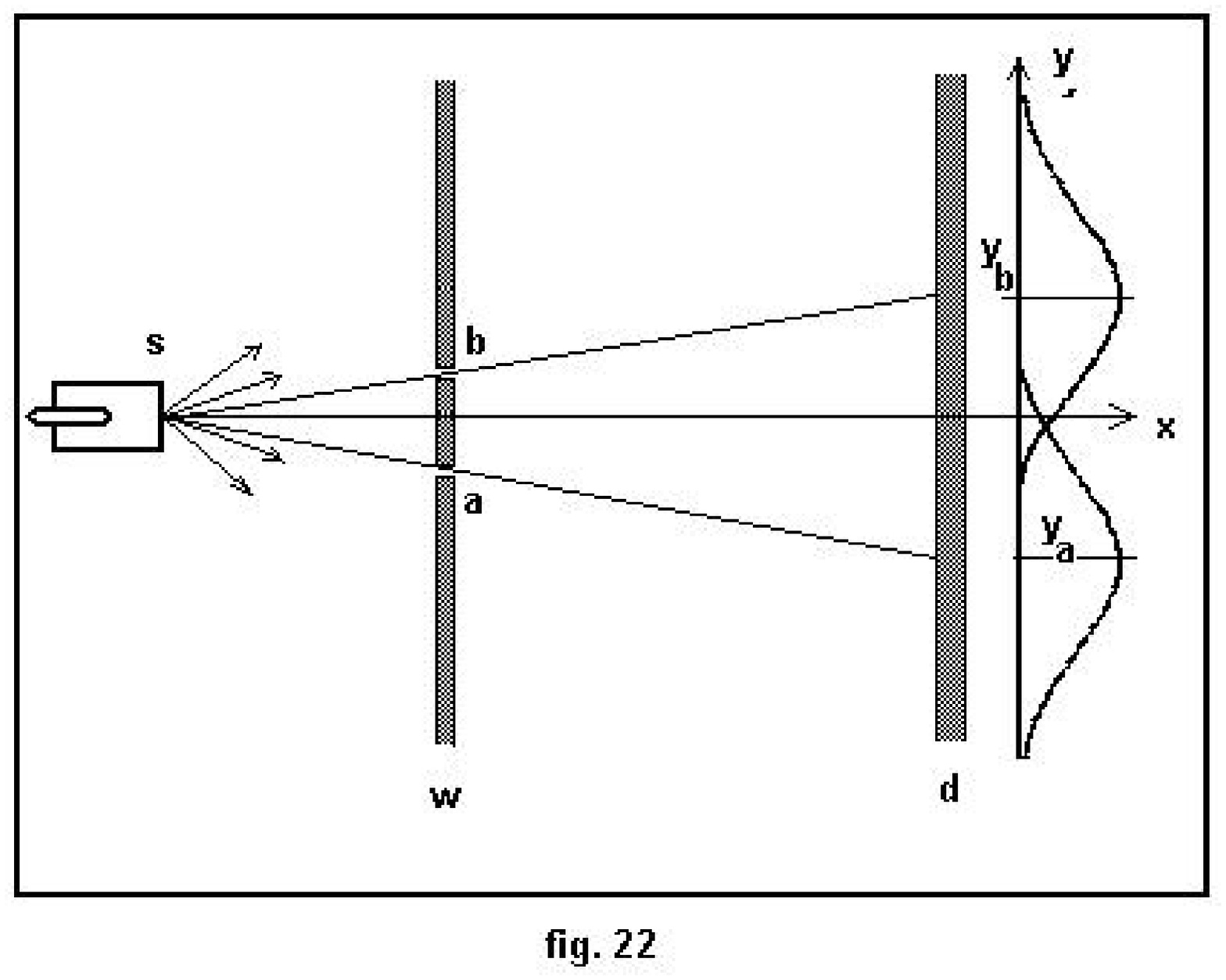}
\caption{}
\label{fig:22}
\end{figure}

\rule{12pt}{0pt} In vacuum (Fig. \ref{fig:20}, Fig. \ref{fig:21}, Fig. \ref{fig:22}): 
Here transmitter $s$ of electrons,
wall $w$ and electron detecting black screen $d$ are placed \cite{Fmn}.

\rule{12pt}{0pt} Electrons are emitted one by one from the source $s$. When an electron hits against screen 
$d$ a bright spot arises in the place of clash on $d$.  

\rule{12pt}{0pt} 1. Let slit $a$ be opened in wall $w$ (Fig. \ref{fig:20}). An electron flies out of $%
s$, passes by $a$, and is detected by $d$.

\rule{12pt}{0pt} If such operation will be reiterated $N$ times then $N$ bright spots
will arise on $d$ against slit $a$ in the vicinity of point $y_a$.

\rule{12pt}{0pt} 2. Let slit $b$ be opened in wall $w$ (Fig. \ref{fig:21}). An electron flies out of $%
s$, passes by $b$, and is detected by $d$.

\rule{12pt}{0pt} If such operation will be reiterated $N$ times then $N$ bright spots
will arise on $d$ against slit $b$ in the vicinity of point $y_b$.

In this case the result like on fig. 3 is expected, isn't it? But no. We get the same result as on

\begin{figure}
\includegraphics[width=0.7\textwidth]{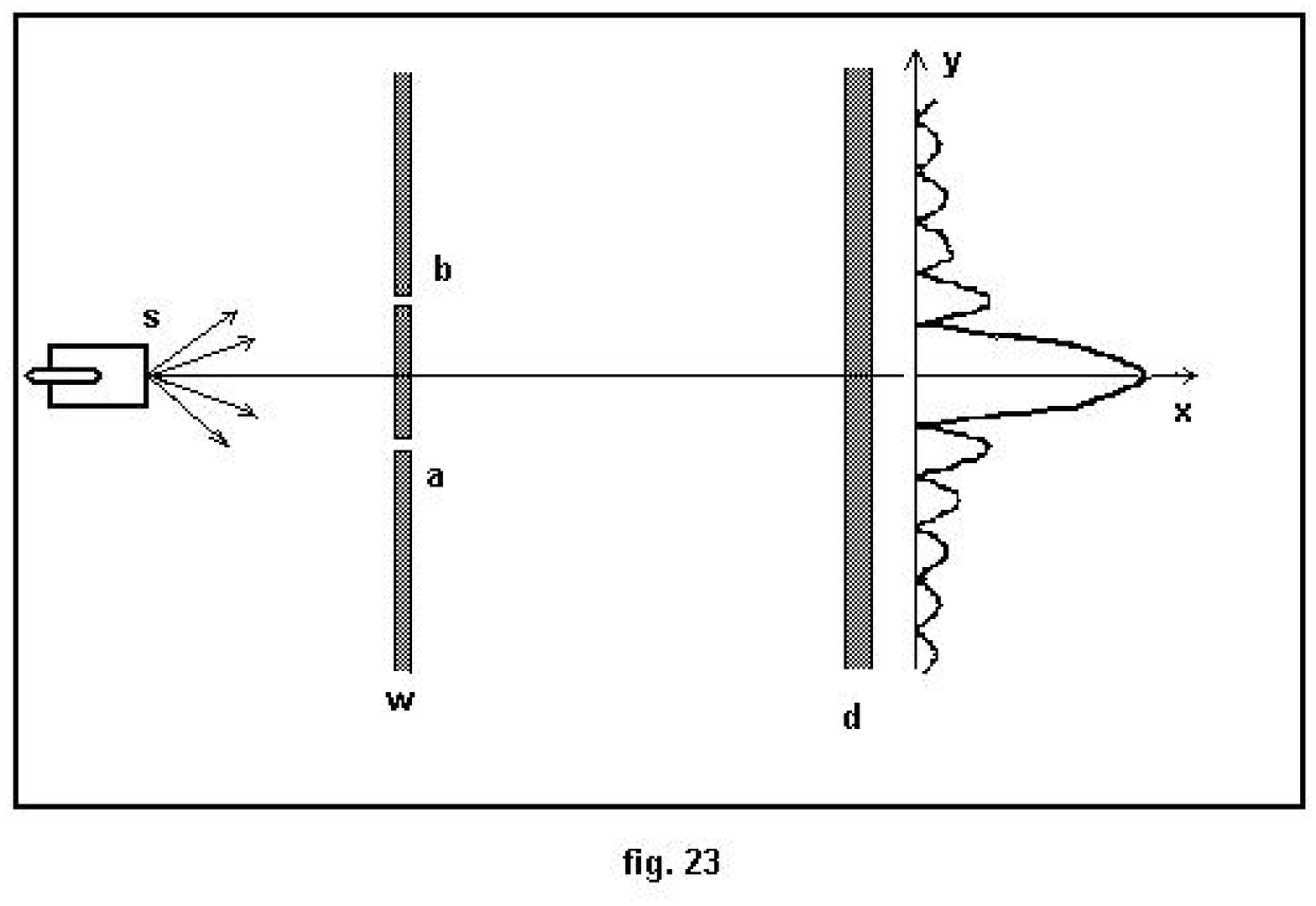}
\caption{}
\label{fig:23}
\end{figure}

\rule{12pt}{0pt} 3. Let both slits be opened. In this case the result like Fig. \ref{fig:22} is expected, 
isn't it? But no. We get the same result as on Fig. \ref{fig:23}\footnote{%
Single-electron events build up over a 20 minute exposure to form an
interference pattern in this double-slit experiment by Akira Tonomura and
co-workers. (a) 8 electrons; (b) 270 electrons; (c) 2000  electrons; (d)
60,000. A video of this experiment will soon be available on the web\\
(www.hqrd.hitachi.co.jp/em/doubleslit.html).}\cite{Ht}.

\begin{figure}
\includegraphics[width=0.7\textwidth]{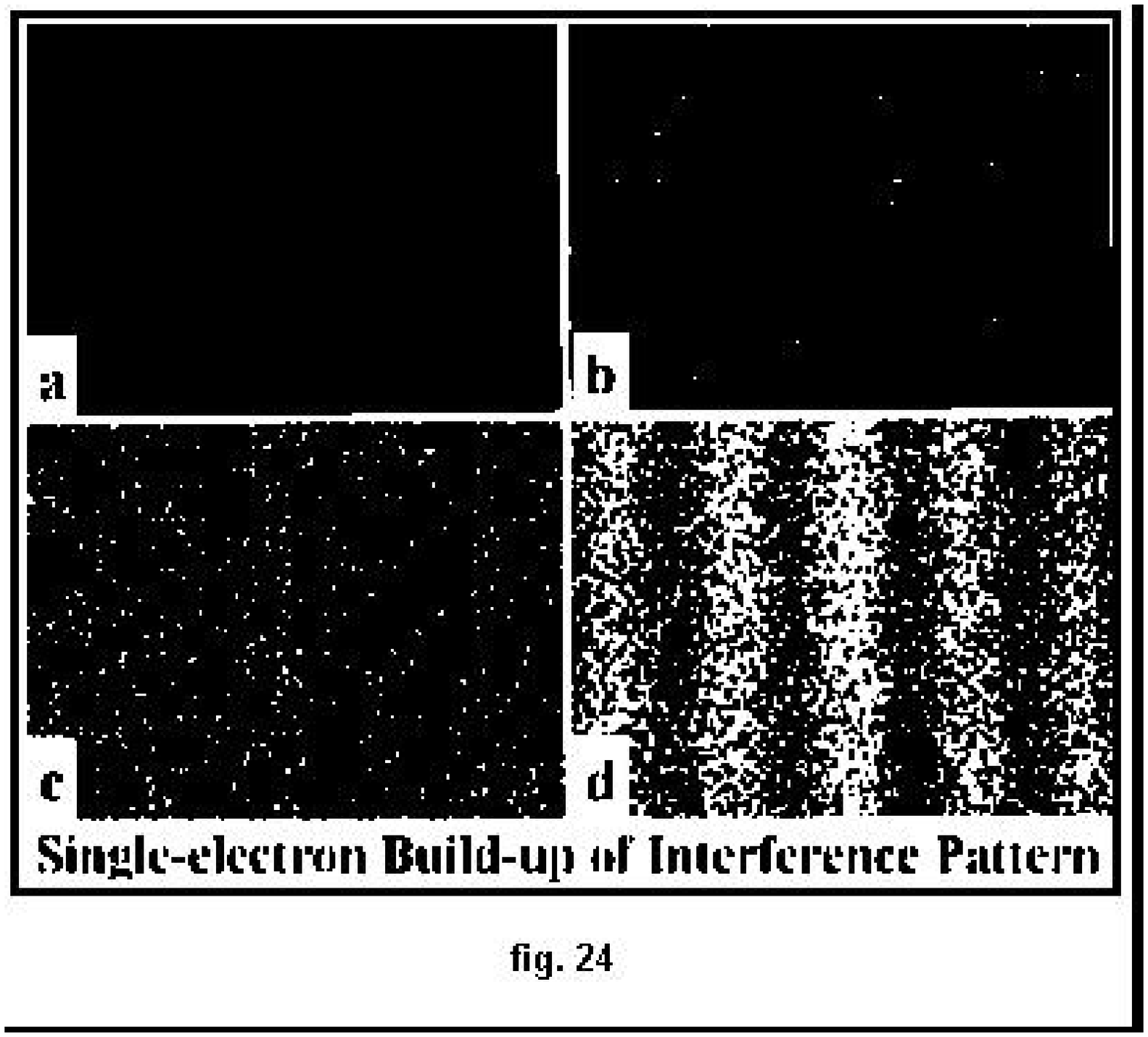}
\caption{}
\label{fig:24}
\end{figure}

\rule{12pt}{0pt} For instance, such experiment was realized at Hitachi by A. Tonomura, J. Endo,
T. Matsuda, T. Kawasaki and H. Ezawa in 1989. It was presumed that interference fringes 
are produced only when two electrons pass through both slits simultaneously. If there
were two electrons from the source $s$ at the same time, such interference
might happen. But this cannot occur, because here is no more than one
electron from this source at one time. Please keep watching the experiment a
little longer. When a large number of electrons is accumulated, something
like regular fringes begin to appear in the perpendicular direction as Fig. \ref{fig:24}(c) 
shows. Clear interference fringes can be seen in the last scene of the
experiment after 20 minutes (Fig. \ref{fig:24}(d)). It should also be noted that the
fringes are made up of bright spots, each of which records the detection of
an electron. We have reached a mysterious conclusion. Although electrons
were sent one by one, interference fringes could be observed. These
interference fringes are formed only when electron waves pass through on
both slits at the same time but nothing other than this. Whenever electrons
are observed, they are always detected as individual particles. When
accumulated, however, interference fringes are formed. Please recall that at
any one instant here was at most one electron from $s$. We have reached a
conclusion which is far from what our common sense tells us.

\rule{12pt}{0pt} 4. But nevertheless, across which slit the electron has slipped? 

\begin{figure}
\includegraphics[width=0.7\textwidth]{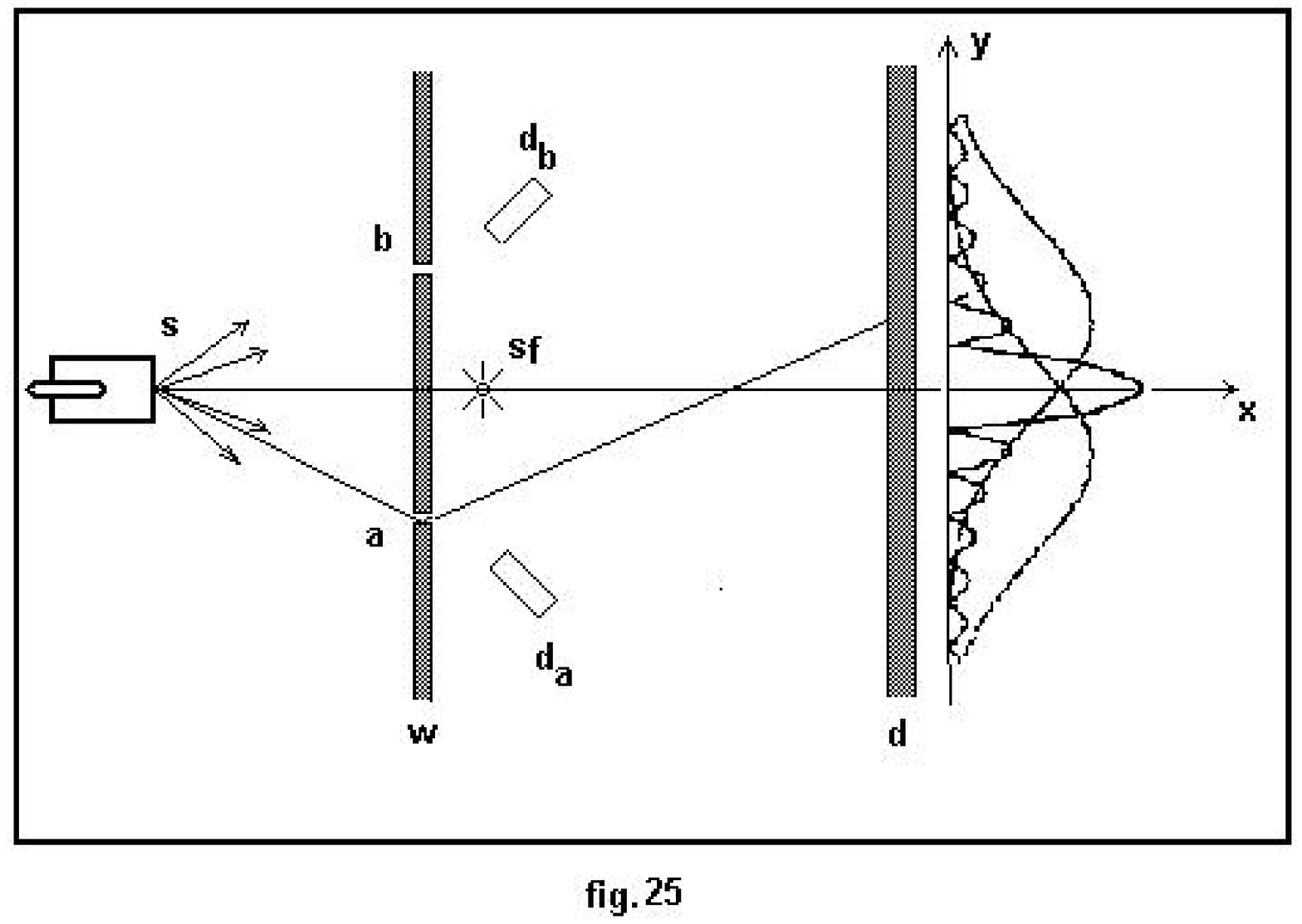}
\caption{}
\label{fig:25}
\end{figure}

\rule{12pt}{0pt} Let (Fig. \ref{fig:25}) two detectors $d_a$ and $d_b$ and a photon source $sf$ be
added to devices of Fig. \ref{fig:23}.

\rule{12pt}{0pt} An electron slipped across slit $a$ is lighten by source $sf$ and
detector $d_a$ snaps into action. And an electron slipped across slit $b$
is lighten by source $sf$ and detector $d_b$ snaps into action.

\rule{12pt}{0pt} If photon source $sf$ lights all $N$ electrons slipped across slits 
we received the picture of Fig. \ref{fig:22}.

\rule{12pt}{0pt} If source $sf$ is faint then only a little part of $N$ electrons slipped
across slits is noticed by detectors $d_a$ and $d_b$. In that case
electrons noticed by detectors $d_a$ and $d_b$ make picture presented on Fig. \ref{fig:22}
and all unnoticed electrons make picture presented on Fig. \ref{fig:23}. In result the 
Fig. \ref{fig:25} is recieved.

\section{Event-Probability Interpretation}

\rule{12pt}{0pt} Let us try to interpret these experiments by events and probabilities.

\rule{12pt}{0pt} Let source $s$ coordinates be $\left\langle x_0,y_0\right\rangle $,
the slit $a$ coordinates be $\left\langle x_a,y_a\right\rangle $, the slit $%
b $ coordinates be $\left\langle x_b,y_b\right\rangle $. Here $x_a=x_b$ and
the wall $w$ equation is $x=x_a$. Let screen $d$ equation be $x=x_d$.

\rule{12pt}{0pt} Denote: \\an event, expressed by sentence: $\ll$electron is detected in point 
$\left\langle t,x,y\right\rangle $$\gg$, as $\mathcal{C}\left( t,x,y\right) $%
,\\an event, expressed by sentence $\ll$slit $a$ is open$\gg$, as $\mathcal{A}$%
, \\and an event, expressed by sentence $\ll$slit $b$ is open$\gg$, as $%
\mathcal{B}$.

\rule{12pt}{0pt} Let $t_0$ be a moment of time when an electron is emitted from source $s$. Since $%
s$ is a pointlike source a state vector $\varphi _{\mathcal{C}}$ in
instant $t_0$ has the following form:

\begin{equation}
\varphi _{\mathcal{C}}\left( t,x,y\right) |_{t=t_0}=\varphi _{\mathcal{C}%
}\left( t_0,x,y\right) \delta \left( x-x_0\right) \delta \left( y-y_0\right) %
\mbox{.}  \label{fii}
\end{equation}

\rule{12pt}{0pt} Let $t_w$ be a moment of time such that if event $\mathcal{C}\left(
t,x,y\right) $ occurs in that instant then $\mathcal{C}\left( t,x,y\right) $
occurs on wall $w$.

\rule{12pt}{0pt} Let $t_d$ be a moment of time of an electron detecting screen $d$.

\rule{12pt}{0pt} 1. Let slit $a$ be opened in wall $w$ (Fig. \ref{fig:20}).

\rule{12pt}{0pt} In that case the $\mathcal{C}\left( t,x,y\right) $ probabilities propagator 

\[
K_{\mathcal{CA}}\left( t-t_0,x-x_s,y-y_s\right) 
\]

in instant $t_w$ should be of the following shape:

\begin{eqnarray*}
&&K_{\mathcal{CA}}\left( t-t_0,x-x_s,y-y_s\right) |_{t=t_w} \\
&=&K_{\mathcal{CA}}\left( t_w-t_0,x-x_s,y-y_s\right) \delta \left(
x-x_a\right) \delta \left( y-y_a\right) \mbox{.}
\end{eqnarray*}

\rule{12pt}{0pt} According to the propagator property:

\begin{eqnarray*}
\  &&K\left( t-t_0,x-x_s,y-y_s\right) = \\
\  &=&\int_{-\frac{\pi \mathrm{c}}{\mathrm{h}}}^{\frac{\pi \mathrm{c}}{%
\mathrm{h}}}dx_1\int_{-\frac{\pi \mathrm{c}}{\mathrm{h}}}^{\frac{\pi \mathrm{%
c}}{\mathrm{h}}}dy_1\cdot K\left( t-t_1,x-x_1,y-y_1\right) \times  \\
&&\times K\left( t_1-t_0,x_1-x_s,y_1-y_s\right) \mbox{.}
\end{eqnarray*}

\rule{12pt}{0pt} Hence,

\begin{eqnarray*}
\  &&K_{\mathcal{CA}}\left( t_d-t_0,x_d-x_s,y_d-y_s\right) = \\
\  &=&\int_{-\frac{\pi \mathrm{c}}{\mathrm{h}}}^{\frac{\pi \mathrm{c}}{%
\mathrm{h}}}dx\int_{-\frac{\pi \mathrm{c}}{\mathrm{h}}}^{\frac{\pi \mathrm{c}%
}{\mathrm{h}}}dy\cdot K_{\mathcal{CA}}\left( t_d-t_w,x_d-x,y_d-y\right)
\times  \\
&&\times K_{\mathcal{CA}}\left( t_w-t_0,x-x_s,y-y_s\right) \delta \left(
x-x_a\right) \delta \left( y-y_a\right) \mbox{.}
\end{eqnarray*}

\rule{12pt}{0pt} Therefore, according to properties of $\delta $-function:

\begin{eqnarray*}
\ &&K_{\mathcal{CA}}\left( t_d-t_0,x_d-x_s,y_d-y_s\right) = \\
\ &=&K_{\mathcal{CA}}\left( t_d-t_w,x_d-x_a,y_d-y_a\right) K_{\mathcal{CA}%
}\left( t_w-t_0,x_a-x_s,y_a-y_s\right) \mbox{.}
\end{eqnarray*}

\rule{12pt}{0pt} The state vector for the event $\mathcal{C}\left( t,x,y\right) $ in
condition $\mathcal{A}$ probability has got the following form (\ref{prp}):

\begin{eqnarray*}
&&\varphi _{\mathcal{CA}}\left( t_d,x_d,y_d\right)=  \\
&=&\int_{-\frac{\pi \mathrm{c}}{\mathrm{h}}}^{\frac{\pi \mathrm{c}}{\mathrm{h%
}}}dx_s\int_{-\frac{\pi \mathrm{c}}{\mathrm{h}}}^{\frac{\pi \mathrm{c}}{%
\mathrm{h}}}dy_s\cdot K_{\mathcal{CA}}\left( t_d-t_0,x_d-x_s,y_d-y_s\right)
\varphi _{\mathcal{C}}\left( t_0,x_s,y_s\right) \mbox{.}
\end{eqnarray*}

\rule{12pt}{0pt} Hence, from (\ref{fii}):

\begin{eqnarray*}
\varphi _{\mathcal{CA}}\left( t_d,x_d,y_d\right) &=&\int_{-\frac{\pi \mathrm{c}}%
{\mathrm{h}}}^{\frac{\pi \mathrm{c}}{\mathrm{h}}} dx_s\int_{-\frac{\pi \mathrm{c}}%
{\mathrm{h}}}^{\frac{\pi \mathrm{c}}{\mathrm{h}}} dy_s\cdot
K_{\mathcal{CA}}\left( t_d-t_0,x_d-x_s,y_d-y_s\right) \\
&&\varphi _{\mathcal{C}}\left( t_0,x_s,y_s\right) \delta \left(
x_s-x_0\right) \delta \left( y_s-y_0\right) \mbox{.}
\end{eqnarray*}

\rule{12pt}{0pt} That is:

\begin{eqnarray*}
&&\varphi _{\mathcal{CA}}\left( t_d,x_d,y_d\right)=  \\
&&=\int\limits_{-\frac{\pi \mathrm{c}}{\mathrm{h}}}^{\frac{\pi \mathrm{c}}{\mathrm{h}}}%
 dx_s\int\limits_{-\frac{\pi \mathrm{c}}{\mathrm{h}}}^{\frac{\pi \mathrm{c}}{\mathrm{h}}}%
dy_s\cdot K_{\mathcal{CA}}\left(
t_d-t_w,x_d-x_a,y_d-y_a\right)  \\
&&\times K_{\mathcal{CA}}\left( t_w-t_0,x_a-x_s,y_a-y_s\right) \times  \\
&&\times \varphi _{\mathcal{C}}\left( t_0,x_s,y_s\right) \delta \left(
x_s-x_0\right) \delta \left( y_s-y_0\right) \mbox{.}
\end{eqnarray*}

\rule{12pt}{0pt} Hence, according to properties of $\delta $-function:

\begin{eqnarray*}                      
&&\varphi _{\mathcal{CA}}\left( t_d,x_d,y_d\right)=  \\
&&=K_{\mathcal{CA}}\left( t_d-t_w,x_d-x_a,y_d-y_a\right)  \\
&&K_{\mathcal{CA}}\left( t_w-t_0,x_a-x_0,y_a-y_0\right)  \\
&&\varphi _{\mathcal{C}}\left( t_0,x_0,y_0\right) \mbox{.}
\end{eqnarray*}

\rule{12pt}{0pt} In accordance with (\ref{j}):

\[
\rho _{\mathcal{CA}}\left( t_d,x_d,y_d\right) =\varphi _{\mathcal{CA}%
}^{\dagger }\left( t_d,x_d,y_d\right) \varphi _{\mathcal{CA}}\left(
t_d,x_d,y_d\right) \mbox{.} 
\]

\rule{12pt}{0pt} Therefore, a probability to detect the electron in vicinity $\Delta x\Delta y
$ of point $\left\langle x_d,y_d\right\rangle $ in instant $t$ in condition $%
\mathcal{A}$ equals to the following:

\[
P_a\left( t_d,x_d,y_d\right) :=\mathrm{P}\left( \mathcal{C}%
\left( t_d,\Delta x\Delta y\right) /\mathcal{A}\right) =\rho _{\mathcal{CA}%
}\left( t_d,x_d,y_d\right) \Delta x\Delta y\mbox{.} 
\]

\rule{12pt}{0pt} 2. Let slit $b$ be opened in wall $w$ (Fig. \ref{fig:21}).

\rule{12pt}{0pt} In that case the $\mathcal{C}\left( t,x,y\right) $ probabilities propagator 

\[
K_{\mathcal{CB}}\left( t-t_0,x-x_s,y-y_s\right) 
\]

in instant $t_w$ should be of the following shape:

\begin{eqnarray*}
&&K_{\mathcal{CB}}\left( t-t_0,x-x_s,y-y_s\right) |_{t=t_w} \\
&=&K_{\mathcal{CB}}\left( t_w-t_0,x-x_s,y-y_s\right) \delta \left(
x-x_b\right) \delta \left( y-y_b\right) \mbox{.}
\end{eqnarray*}

\rule{12pt}{0pt} Hence, according to the propagator property:

\begin{eqnarray*}
\ &&K_{\mathcal{CB}}\left( t_d-t_0,x_d-x_s,y_d-y_s\right) = \\
\ &=&\int\limits_{-\frac{\pi \mathrm{c}}{\mathrm{h}}}^{\frac{\pi \mathrm{c}}{\mathrm{h}}}%
dx\int\limits_{-\frac{\pi \mathrm{c}}{\mathrm{h}}}^{\frac{\pi \mathrm{c}}{\mathrm{h}}}dy%
\cdot K_{\mathcal{CB}}\left( t_d-t_w,x_d-x,y_d-y\right)
\\
&&K_{\mathcal{CB}}\left( t_w-t_0,x-x_s,y-y_s\right) \delta \left(
x-x_b\right) \delta \left( y-y_b\right) \mbox{.}
\end{eqnarray*}

\rule{12pt}{0pt} Therefore, according to properties of $\delta $-function:

\begin{eqnarray*}
\ &&K_{\mathcal{CB}}\left( t_d-t_0,x_d-x_s,y_d-y_s\right) = \\
\ &=&K_{\mathcal{CB}}\left( t_d-t_w,x_d-x_b,y_d-y_b\right) K_{\mathcal{CB}%
}\left( t_w-t_0,x_b-x_s,y_b-y_s\right) \mbox{.}
\end{eqnarray*}

\rule{12pt}{0pt} The state vector for the event $\mathcal{C}\left( t,x,y\right) $ in
condition $\mathcal{B}$ probability has got the following form (\ref{prp}):

\begin{eqnarray*}
&&\varphi _{\mathcal{CB}}\left( t_d,x_d,y_d\right)=  \\
&=&\int_{-\frac{\pi \mathrm{c}}{\mathrm{h}}}^{\frac{\pi \mathrm{c}}{\mathrm{h%
}}}dx_s\int_{-\frac{\pi \mathrm{c}}{\mathrm{h}}}^{\frac{\pi \mathrm{c}}{%
\mathrm{h}}}dy_s\cdot K_{\mathcal{CB}}\left( t_d-t_0,x_d-x_s,y_d-y_s\right)
\varphi _{\mathcal{C}}\left( t_0,x_s,y_s\right) \mbox{.}
\end{eqnarray*}

\rule{12pt}{0pt} Hence, from (\ref{fii}):

\begin{eqnarray*}
\varphi _{\mathcal{CB}}\left( t_d,x_d,y_d\right) &=&\int_{-\frac{\pi \mathrm{c}}%
{\mathrm{h}}}^{\frac{\pi \mathrm{c}}{\mathrm{h}}} dx_s\int_{-\frac{\pi \mathrm{c}}%
{\mathrm{h}}}^{\frac{\pi \mathrm{c}}{\mathrm{h}}} dy_s\cdot
K_{\mathcal{CB}}\left( t_d-t_0,x_d-x_s,y_d-y_s\right) \\
&&\times\varphi _{\mathcal{C}}\left( t_0,x_s,y_s\right) \delta \left(
x_s-x_0\right) \delta \left( y_s-y_0\right) \mbox{.}
\end{eqnarray*}

\rule{12pt}{0pt} That is:

\begin{eqnarray*}
&&\varphi _{\mathcal{CB}}\left( t_d,x_d,y_d\right)=  \\
&&=\int\limits_{-\frac{\pi \mathrm{c}}{\mathrm{h}}}^{\frac{\pi \mathrm{c}}{\mathrm{h}}}%
dx_s\int\limits_{-\frac{\pi \mathrm{c}}{\mathrm{h}}}^{\frac{\pi \mathrm{c}}{\mathrm{h}}}%
dy_s\cdot K_{\mathcal{CB}}\left(
t_d-t_w,x_d-x_b,y_d-y_b\right)  \\
&&\times K_{\mathcal{CB}}\left( t_w-t_0,x_b-x_s,y_b-y_s\right)  \\
&&\times \varphi _{\mathcal{C}}\left( t_0,x_s,y_s\right) \delta \left(
x_s-x_0\right) \delta \left( y_s-y_0\right) \mbox{.}
\end{eqnarray*}

\rule{12pt}{0pt} Hence, according to properties of $\delta $-function:

\begin{eqnarray*}
&&\varphi _{\mathcal{CB}}\left( t_d,x_d,y_d\right)=  \\
&&=K_{\mathcal{CB}}\left( t_d-t_w,x_d-x_b,y_d-y_b\right) \times  \\
&&K_{\mathcal{CB}}\left( t_w-t_0,x_b-x_0,y_b-y_0\right) \times  \\
&&\varphi _{\mathcal{C}}\left( t_0,x_0,y_0\right) \mbox{.}
\end{eqnarray*}

\rule{12pt}{0pt} In accordance with (\ref{j}):

\[
\rho _{\mathcal{CB}}\left( t_d,x_d,y_d\right) =\varphi _{\mathcal{CB}%
}^{\dagger }\left( t_d,x_d,y_d\right) \varphi _{\mathcal{CB}}\left(
t_d,x_d,y_d\right) \mbox{.} 
\]

\rule{12pt}{0pt} Therefore, a probability to detect the electron in vicinity $\Delta x\Delta y
$ of point $\left\langle x_d,y_d\right\rangle $ in instant $t$ in condition $%
\mathcal{B}$ equals to the following:

\[
P_b\left( t_d,x_d,y_d\right) :=\mathrm{P}\left( \mathcal{C}%
\left( t_d,\Delta x\Delta y\right) /\mathcal{B}\right) =\rho _{\mathcal{CB}%
}\left( t_d,x_d,y_d\right) \Delta x\Delta y \mbox{.} 
\]

\rule{12pt}{0pt} 3. Let both slits and $a$ and $b$ are opened (Fig. \ref{fig:23}).

\rule{12pt}{0pt} In that case the $\mathcal{C}\left( t,x,y\right) $ probabilities propagator
 
\[
K_{\mathcal{CAB}}\left( t-t_0,x-x_s,y-y_s\right) 
\]

in instant $t_w$ should be of the following shape:

\[
\begin{array}{c}
\ K_{\mathcal{CAB}}\left( t-t_0,x-x_s,y-y_s\right) |_{t=t_w}= \\ 
=K_{\mathcal{CAB}}\left( t_w-t_0,x-x_s,y-y_s\right) \times  \\ 
\left( \delta \left( x-x_a\right) \delta \left( y-y_a\right) +\delta \left(
x-x_b\right) \delta \left( y-y_b\right) \right) \mbox{.}
\end{array}
\]

\rule{12pt}{0pt} Hence, according to the propagator property:

\[
\begin{array}{c}
\ K_{\mathcal{CAB}}\left( t_d-t_0,x_d-x_s,y_d-y_s\right) = \\ 
\ =\int\limits_{-\frac{\pi \mathrm{c}}{\mathrm{h}}}^{\frac{\pi \mathrm{c}}{\mathrm{h}}}%
dx\int\limits_{-\frac{\pi \mathrm{c}}{\mathrm{h}}}^{\frac{\pi \mathrm{c}}{\mathrm{h}}}%
dy\cdot K_{\mathcal{CAB}}\left( t_d-t_w,x_d-x,y_d-y\right)\\ 
\times K_{\mathcal{CAB}}\left( t_w-t_0,x-x_s,y-y_s\right) \\ 
\times\left( \delta \left( x-x_a\right) \delta \left( y-y_a\right) +\delta
\left( x-x_b\right) \delta \left( y-y_b\right) \right) \mbox{.}
\end{array}
\]

\rule{12pt}{0pt} Hence,

\[
\begin{array}{c}
\ K_{\mathcal{CAB}}\left( t_d-t_0,x_d-x_s,y_d-y_s\right) = \\ 
\begin{array}{c}
\int\limits_{-\frac{\pi \mathrm{c}}{\mathrm{h}}}^{\frac{\pi \mathrm{c}}{\mathrm{h}}}%
dx\int\limits_{-\frac{\pi \mathrm{c}}{\mathrm{h}}}^{\frac{\pi \mathrm{c}}{\mathrm{h}}}%
dy\cdot K_{\mathcal{CAB}}\left( t_d-t_w,x_d-x,y_d-y\right) \\ 
\times K_{\mathcal{CAB}}\left( t_w-t_0,x-x_s,y-y_s\right) \times  \\ 
\times \delta \left( x-x_a\right) \delta \left( y-y_a\right)  \\ 
+\int\limits_{-\frac{\pi \mathrm{c}}{\mathrm{h}}}^{\frac{\pi \mathrm{c}}{\mathrm{h}}}%
dx\int\limits_{-\frac{\pi \mathrm{c}}{\mathrm{h}}}^{\frac{\pi \mathrm{c}}{\mathrm{h}}}%
dy\cdot K_{\mathcal{CAB}}\left( t_d-t_w,x_d-x,y_d-y\right) \\ 
\times K_{\mathcal{CAB}}\left( t_w-t_0,x-x_s,y-y_s\right) \times  \\ 
\times \delta \left( x-x_b\right) \delta \left( y-y_b\right) \mbox{.}
\end{array}
\end{array}
\]

\rule{12pt}{0pt} Hence, according to properties of $\delta $-function:

\[
\begin{array}{c}
\ K_{\mathcal{CAB}}\left( t_d-t_0,x_d-x_s,y_d-y_s\right) = \\ 
\begin{array}{c}
K_{\mathcal{CAB}}\left( t_d-t_w,x_d-x_a,y_d-y_a\right) \ K_{\mathcal{CAB}%
}\left( t_w-t_0,x_a-x_s,y_a-y_s\right) \\ 
+K_{\mathcal{CAB}}\left( t_d-t_w,x_d-x_b,y_d-y_b\right) \ K_{\mathcal{CAB}%
}\left( t_w-t_0,x_b-x_s,y_b-y_s\right)
\end{array}
\mbox{.}
\end{array}
\]

\rule{12pt}{0pt} The state vector for the event $\mathcal{C}\left( t,x,y\right) $ in
condition $\mathcal{A}$ and $\mathcal{B}$ probability has the following
form (\ref{prp}):

\begin{eqnarray*}
&&\varphi _{\mathcal{CAB}}\left( t_d,x_d,y_d\right)=  \\
&&=\int\limits_{-\frac{\pi \mathrm{c}}{\mathrm{h}}}^{\frac{\pi \mathrm{c}}{\mathrm{h}}}%
dx_s\int\limits_{-\frac{\pi \mathrm{c}}{\mathrm{h}}}^{\frac{\pi \mathrm{c}}{\mathrm{h}}}%
dy_s\cdot K_{\mathcal{CAB}}\left(
t_d-t_0,x_d-x_s,y_d-y_s\right) \varphi _{\mathcal{C}}\left(
t_0,x_s,y_s\right) \mbox{.}
\end{eqnarray*}

\rule{12pt}{0pt} Hence, from (\ref{fii}):

\begin{eqnarray*}
\varphi _{\mathcal{CAB}}\left( t_d,x_d,y_d\right) &=&\int_{-\frac{\pi \mathrm{c}}%
{\mathrm{h}}}^{\frac{\pi \mathrm{c}}{\mathrm{h}}} dx_s\int_{-\frac{\pi \mathrm{c}}%
{\mathrm{h}}}^{\frac{\pi \mathrm{c}}{\mathrm{h}}} dy_s\cdot
K_{\mathcal{CAB}}\left( t_d-t_0,x_d-x_s,y_d-y_s\right) \\
&&\times \varphi _{\mathcal{C}}\left( t_0,x_s,y_s\right) \delta \left(
x_s-x_0\right) \delta \left( y_s-y_0\right) \mbox{.}
\end{eqnarray*}

\rule{12pt}{0pt} That is:

\[
\begin{array}{c}
\varphi _{\mathcal{CAB}}\left( t_d,x_d,y_d\right) =\int_{-\frac{\pi \mathrm{c}}%
{\mathrm{h}}}^{\frac{\pi \mathrm{c}}{\mathrm{h}}} dx_s\int_{-\frac{\pi \mathrm{c}}%
{\mathrm{h}}}^{\frac{\pi \mathrm{c}}{\mathrm{h}}} dy_s
\\ 
\times \left( 
\begin{array}{c}
K_{\mathcal{CAB}}\left( t_d-t_w,x_d-x_a,y_d-y_a\right) \ K_{\mathcal{CAB}%
}\left( t_w-t_0,x_a-x_s,y_a-y_s\right) \\ 
+K_{\mathcal{CAB}}\left( t_d-t_w,x_d-x_b,y_d-y_b\right) \ K_{\mathcal{CAB}%
}\left( t_w-t_0,x_b-x_s,y_b-y_s\right)
\end{array}
\right) \\ 
\times \varphi _{\mathcal{C}}\left( t_0,x_s,y_s\right) \delta \left(
x_s-x_0\right) \delta \left( y_s-y_0\right) \mbox{.}
\end{array}
\]

\rule{12pt}{0pt} Hence, according to properties of $\delta $-function:

\[
\begin{array}{c}
\ \varphi _{\mathcal{CAB}}\left( t_d,x_d,y_d\right) = \\ 
=\left( 
\begin{array}{c}
K_{\mathcal{CAB}}\left( t_d-t_w,x_d-x_a,y_d-y_a\right) \ K_{\mathcal{CAB}%
}\left( t_w-t_0,x_a-x_0,y_a-y_0\right) \\ 
+K_{\mathcal{CAB}}\left( t_d-t_w,x_d-x_b,y_d-y_b\right) \ K_{\mathcal{CAB}%
}\left( t_w-t_0,x_b-x_0,y_b-y_0\right)
\end{array}
\right) \\ 
\times\varphi _{\mathcal{C}}\left( t_0,x_0,y_0\right) \mbox{.}
\end{array}
\]

\rule{12pt}{0pt} That is:

\[
\begin{array}{c}
\ \varphi _{\mathcal{CAB}}\left( t_d,x_d,y_d\right) = \\ 
=K_{\mathcal{CAB}}\left( t_d-t_w,x_d-x_a,y_d-y_a\right) \ \times  \\ 
K_{\mathcal{CAB}}\left( t_w-t_0,x_a-x_0,y_a-y_0\right) \varphi _{\mathcal{C}%
}\left( t_0,x_0,y_0\right)  \\ 
+K_{\mathcal{CAB}}\left( t_d-t_w,x_d-x_b,y_d-y_b\right) \ \times  \\ 
K_{\mathcal{CAB}}\left( t_w-t_0,x_b-x_0,y_b-y_0\right)  \\ 
\times \varphi _{\mathcal{C}}\left( t_0,x_0,y_0\right) \mbox{.}
\end{array}
\]

\rule{12pt}{0pt} Therefore,

\[
\varphi _{\mathcal{CAB}}\left( t_d,x_d,y_d\right) =\varphi_{\mathcal{CA}%
}\left( t_d,x_d,y_d\right) +\varphi_{\mathcal{CB}}\left( t_d,x_d,y_d\right) %
\mbox{.} 
\]

\rule{12pt}{0pt} And in accordance with (\ref{j}):

\[
\rho _{\mathcal{CAB}}\left( t_d,x_d,y_d\right) =\varphi _{\mathcal{CAB}%
}^{\dagger }\left( t_d,x_d,y_d\right) \varphi _{\mathcal{CAB}}\left(
t_d,x_d,y_d\right) \mbox{,} 
\]

i.e.

\[
\rho _{\mathcal{CAB}}=\left( \varphi _{\mathcal{CA}}+\varphi _{\mathcal{CB}%
}\right) ^{\dagger }\left( \varphi _{\mathcal{CA}}+\varphi _{\mathcal{CB}%
}\right) 
\]

\rule{12pt}{0pt} Since state vectors $\varphi _{\mathcal{CA}}$ and $\varphi _{\mathcal{CB}}$
are not numbers with like signs then in the general case:

\[
\left( \varphi _{\mathcal{CA}}+\varphi _{\mathcal{CB}}\right) ^{\dagger
}\left( \varphi _{\mathcal{CA}}+\varphi _{\mathcal{CB}}\right) \neq \varphi
_{\mathcal{CA}}^{\dagger }\varphi _{\mathcal{CA}}+\varphi _{\mathcal{CB}%
}^{\dagger }\varphi _{\mathcal{CB}}\mbox{.} 
\]

\rule{12pt}{0pt} Therefore, since a probability to detect the electron in vicinity $\Delta
x\Delta y$ of point $\left\langle x_d,y_d\right\rangle $ in instant $t$ in
condition $\mathcal{AB}$ equals:

\[
P_{ab}\left( t_d,x_d,y_d\right) :=\mathrm{P}\left( \mathcal{C}%
\left( t_d,\Delta x\Delta y\right) /\mathcal{AB}\right) =\rho _{\mathcal{CAB}%
}\left( t_d,x_d,y_d\right)\Delta x\Delta y 
\]

then

\[
P_{ab}\left( t_d,x_d,y_d\right) \neq P_a\left( t_d,x_d,y_d\right) +P_b\left(
t_d,x_d,y_d\right) \mbox{.} 
\]

\rule{12pt}{0pt} Hence, we have the Fig. \ref{fig:23} picture instead of the Fig. \ref{fig:22} picture.

\rule{12pt}{0pt} 4. Let us consider devices on Fig. \ref{fig:25}.

\rule{12pt}{0pt} Denote:\\ 
event expressed by sentence ''detector $d_a$ snaps into action'' as 
$\mathcal{D}_a$\\ 
event expressed by sentence ''detector $d_b$ snaps into
action'' as $\mathcal{D}_b$.

\rule{12pt}{0pt} Since event $\mathcal{C}\left( t,x,y\right) $
is a pointlike event then events $\mathcal{D}_a$ and $\mathcal{D}_b$ are
exclusive events.

\rule{12pt}{0pt} According to the property of operations on events:

\[
\left( \mathcal{D}_a+\mathcal{D}_b\right) +\overline{\left( \mathcal{D}_a+%
\mathcal{D}_b\right) }=\mathcal{T} 
\]

where $\mathcal{T}$ is the sure event, and

\[
\overline{\left( \mathcal{D}_a+\mathcal{D}_b\right) }=\overline{\mathcal{D}}%
_a\overline{\mathcal{D}}_b\mbox{,} 
\]

\rule{12pt}{0pt} Hence,

\[
\mathcal{D}_a+\mathcal{D}_b+\overline{\mathcal{D}}_a\overline{\mathcal{D}}_b=%
\mathcal{T}\mbox{.} 
\]

\rule{12pt}{0pt} Hence,

\[
\mathcal{C}=\mathcal{CT}=\mathcal{C}\left( \mathcal{D}_a+\mathcal{D}_b+%
\overline{\mathcal{D}}_a\overline{\mathcal{D}}_b\right) \mbox{.} 
\]

\rule{12pt}{0pt} Hence,

\[
\mathcal{C}=\mathcal{CD}_a+\mathcal{CD}_b+\mathcal{C}\overline{\mathcal{D}}_a%
\overline{\mathcal{D}}_b\mbox{.} 
\]

\rule{12pt}{0pt} Therefore, according to the probabilities addition formula for exclusive events:

\[
\mathrm{P}\left( \mathcal{C}\left( t_d\right) \right) =\mathrm{P}\left( 
\mathcal{C}\left( t_d\right) \mathcal{D}_a\right) +\mathrm{P}\left( \mathcal{%
C}\left( t_d\right) \mathcal{D}_b\right) +\mathrm{P}\left( \mathcal{C}\left(
t_d\right) \overline{\mathcal{D}}_a\overline{\mathcal{D}}_b\right) \mbox{.} 
\]

\rule{12pt}{0pt} But

\begin{eqnarray*}
\mathrm{P}\left( \mathcal{C}\left( t_d\right) \mathcal{D}_a\right)
&=&P_a\left( t_d\right) \mbox{,} \\
\mathrm{P}\left( \mathcal{C}\left( t_d\right) \mathcal{D}_b\right)
&=&P_b\left( t_d\right) \mbox{,} \\
\mathrm{P}\left( \mathcal{C}\left( t_d\right) \overline{\mathcal{D}}_a%
\overline{\mathcal{D}}_b\right) &=&P_{ab}\left( t_d\right) \mbox{,}
\end{eqnarray*}

and we receive the Fig. \ref{fig:25} picture.

\rule{12pt}{0pt} Thus, here are no paradoxes for the event-probability interpretation of these
experiments. We should depart from notion of a continuously existing
electron and consider an elementary particle an ensemble of events 
connected by probability. It's like the fact that physical particle exists
only at the instant when it is involved in some event. A particle doesn't
exist in any other time, but there's a probability that something will
happen to it. Thus, if nothing happens with the particle between the event of creating it and the event of 
detectiting it the behavior of the particle is the behavior of probability between the point 
of creating and the point of detecting it with the presence of interference.

\rule{12pt}{0pt} But what is with Wilson cloud chamber where the particle has a clear trajectory 
and no interference?

\rule{12pt}{0pt} In that case these trajectories are not  totally continuous lines. Every point of 
ionization has neighbouring point of ionization, and there are no events between these
points.
 
\rule{12pt}{0pt} Consequently, physical particle is moving because corresponding probability
propagates in the space between points of ionization.
Consequently, particle is an ensemble of events, connected by probability.
And charges, masses, moments, etc. represent statistical parameters of these probability
waves, propagated in the space-time. It explains all paradoxes
of quantum physics.\linebreak Schrodinger's cat lives easy without any superposition
of states until the microevent awaited by all occures. And the wave function disappears without 
any collapse in the moment when an event probability disappears after the event occurs.
 
\section{Conclusion}

\rule{12pt}{0pt} The Quantum Theory equations describe the behaviour of probabilities of pointlike events.

Double-Slit experiment demonstrates that an elementary particle is an ensemble of such 
events connected by these probabilities.

Quantum Theory is one of the possible ways of processing of probabilistic information.

\end{document}